\documentstyle[12pt]{article}

\begin{document}
\sloppy
\begin{flushright}{SIT-HEP/TM-1\\ February/ 2000}
\end{flushright}
\vskip 1.5 truecm
\centerline{\large{\bf Weak Scale Inflation and Unstable 
Domain Walls}}
\vskip .75 truecm
\centerline{\bf Tomohiro Matsuda
\footnote{matsuda@sit.ac.jp}}
\vskip .4 truecm
\centerline {\it Laboratory of Physics, Saitama Institute of
 Technology,}
\centerline {\it Fusaiji, Okabe-machi, Saitama 369-0293, 
Japan}
\vskip 1. truecm
\makeatletter
\@addtoreset{equation}{section}
\def\theequation{\thesection.\arabic{equation}}
\makeatother
\vskip 1. truecm
\begin{abstract}
\hspace*{\parindent}
Cosmlogical domain walls produced during phase transition
are expected to collapse on a cosmologically safe 
timescale if Vilenkin's condition is satisfied.
 We show that the 
decaying processes of these unstable domain walls should 
be changed significantly if weak scale inflation takes place.
The usual condition for the safe decay of the cosmological
domain wall must be changed, depending on their scales and
interactions.
As a result, the energy scales and explicit breaking
terms of such walls must satisfy severe requirements.
 We also make a brief comment on 
cosmological structure formation.
\end{abstract}

\newpage
\section{Introduction}
\hspace*{\parindent}
Many types of discrete symmetries appear in supersymmetric
models.
They are usually broken at intermediate scales
and cause  the cosmological domain wall problem,
if the walls remain stable.
The usual assumption is that non-renormalizable terms
induced by gravitational interactions may explicitly 
break these symmetries
and make such walls collapse on a cosmologically safe
timescale\cite{vilenkin}.

On the other hand, 
many models of supersymmetry breaking involve particles
with the masses of order the gravitino mass $m_{3/2}$ 
and Planck mass 
suppressed couplings.
Coherent production of such particles in the early universe
destroys the successful predictions of nucleosynthesis.
This problem may be solved by a brief period of weak scale
inflation.

In this paper we examine whether the  usual criterion for
the safe decay of the unstable domain walls can be applied
when weak scale inflation takes place.
In section 2 and 3 we review the usual condition for
unstable domain walls and the basic idea of 
thermal inflation.
In section 4 we show collective examples of discrete
symmetries which appear in many supersymmetric models.
In section 5 
we examine whether the usual criterion for the 
safe decay of the cosmological domain wall is also 
applicable for supersymmetric models.
We show that the decaying process of these unstable domain
walls should be changed significantly if thermal inflation
occurs.
As a result, the scenario for  the safe decay of 
the cosmological domain walls must be changed,
 depending on their scales and  interactions.

We also make a brief comment on the cosmological structure 
formation which will be induced by the soft domain walls.
If such walls are expanded during weak scale inflation,
the conditions for the structure formation will be
 changed.

\section{Collision of the Cosmological Domain Walls}
\hspace*{\parindent}
In this section we briefly review how to estimate the
value of the pressure (i.e., explicit breaking) 
to safely remove the walls.
The crudest estimate we can make is to insist
that the walls are removed before they dominate over
the radiation energy density in the universe.
When the discrete symmetry is broken  by 
gravitational interactions, the symmetry is an approximate
discrete symmetry.
The degeneracy is broken and the energy difference 
$\epsilon\ne0$ appears.
Regions of higher density vacuum tend to shrink,
the corresponding force per unit area of the wall 
is $\sim \epsilon$.
The energy difference $\epsilon$ becomes dynamically 
important
when this force becomes comparable to the force of 
the tension $f\sim \sigma/R$,
where $\sigma$ is the surface energy density of the wall.
For walls to disappear, this has to happen before the
walls dominate the universe.
On the other hand, the domain wall network is not a static 
system.
In general, initial shape of the walls right after the
phase transition is determined by the random variation
of the scalar VEV.
One expects the walls to be very irregular, random
surfaces with a typical curvature radius, which
is determined by the correlation length of the
scalar field.
To characterize the system of domain walls,
one can use a simulation\cite{simulation}.
The system will be dominated by one large (infinite size)
wall network  and some finite closed walls (cells) when
they form.
The isolated closed walls smaller than the horizon
will shrink and disappear soon after the phase transition.
Since the walls smaller than the horizon size 
will efficiently disappear so that only walls
at the horizon size will remain, 
their typical curvature scale will be  the horizon 
size, $R\sim t\sim M_{p}/g_{*}^{\frac{1}{2}}T^{2}$.
Since the energy density of the wall $\rho_{w}$ is about
\begin{equation}
\rho_{w}\sim \frac{\sigma}{R},
\end{equation}
and the radiation energy density $\rho_{r}$ is 
\begin{equation}
\rho_{r}\sim g_{*}T^{4},
\end{equation}
one sees that the wall dominates the evolution 
below a temperature $T_{w}$
\begin{equation}
T_{w}\sim \left(\frac{\sigma}{g_{*}^{1/2}M_{p}}
\right)^{\frac{1}{2}}.
\end{equation}
To prevent the wall domination, one requires the
pressure to have become dominant before this epoch,
\begin{equation}
\label{criterion}
\epsilon>\frac{\sigma}{R_{w}}\sim
\frac{\sigma^{2}}{M^{2}_{p}}.
\end{equation}
Here $R_{w}$ denotes the horizon size at the wall domination.
A pressure of this magnitude would be produced by
higher dimensional operators which explicitly break
 the discrete symmetry.

The criterion (\ref{criterion})
seems appropriate, if the scale of the wall is higher 
than $(10^{5}GeV)^{3}$.
For the walls below this scale ($\sigma\le(10^{5}GeV)^{3}$),
 there should
be  further constraints coming from primordial 
nucleosynthesis.
Since the time associated with the collapsing temperature
 $T_{w}$
is $t_{w}\sim M_{p}^{2}/g_{*}^{\frac{1}{2}}\sigma
\sim 10^{8}\left(\frac{(10^{2}GeV)^{3}}{\sigma}\right)$sec,
the walls $\sigma\le(10^{5}GeV)^{3}$ will decay after 
nucleosynthesis\cite{Abel}.
If the walls are not hidden and can decay into the standard
model particles, the entropy produced when walls collide
will violate the phenomenological bounds for nucleosynthesis.
On the other hand, this simple bound
($\sigma\ge(10^{5}GeV)^{3}$) is not effective 
for the walls which cannot decay into standard model 
particles.
The  walls such as soft domain walls\cite{soft_wall,bias},
the succeeding story should strongly depend on the 
details of the hidden components and their interactions.
These walls can decay late to contribute to the large scale
structure formation.

\section{Weak scale inflation}
\hspace*{\parindent}
Supersymmetry is probably one of the most attractive 
extensions of the standard model.
In virtue of  supersymmetry, the hierarchy can be 
stabilized against the radiative corrections.

However, overviewing the cosmology of the supersymmetric
models, one faces with various difficulties.
One of the most obvious and famous problems is the
gravitino problem\cite{gravitino}.
This problem still exists even if the universe
experiences a primordial inflation, since the gravitino
is reproduced during the reheating process.
The mass of the gravitino depends on the mechanisms of
supersymmetry breaking.
In the supergravity mediated models of supersymmetry
breaking, the gravitino has 
a mass of the electroweak scale ($m_{3/2}\sim10^{2-3}$GeV),
and it decays soon after big bang nucleosynthesis.
High energy photons produced by the gravitino decay may
destroy the usual assumptions for big bang 
nucleosynthesis.
Another example is the gauge mediated supersymmetry breaking
models, in which the predicted gravitino mass is much
lighter than the supergravity mediated models.
The gravitino mass is expected to be $m_{3/2}\sim10$eV$-1$GeV
and cosmologically stable.
If the gravitino mass is larger than $1$keV, the universe
will be overclosed unless the gravitino is diluted
at an earlier epoch.

The gravitino problem is a common feature of the superstring
inspired models, because the moduli fields should play the
same role as the gravitino.
However, the cosmological moduli problem has a
different feature.
Because a moduli is a scalar field, it should have the
potential which is inevitably flat but  raised by
the moduli mass $m_{\phi}\sim m_{3/2}$.
In the supergravity mediated models of supersymmetry 
breaking, the moduli fields decay soon
after big bang nucleosynthesis as the gravitino, 
causing to the same cosmological problem.
For the gauge mediated models which predicts lighter
mass for the moduli, the energy of the oscillation
lasts and overcloses the universe for $m_{3/2}<100$MeV,
 or the decay of the moduli gives too much contribution
 to the x($\gamma$)-ray background spectrum for
$m_{3/2}\sim 10^{-1}-10^{4}$MeV.

The most promising way to evade these difficulties is to
dilute these unwanted relics after the primordial
inflation.
One of such dilution mechanisms is the thermal inflation
model proposed by Lyth and Stewart\cite{thermal}.
Thermal inflation occurs before the electroweak
phase transition and produces entropy before
big bang nucleosynthesis, which dilutes the moduli
density.
During thermal inflation the flaton field (i.e., the
inflation field for thermal inflation) is held at
the origin by finite temperature effects.
The potential energy during thermal inflation is the value
$V_{0}$ of the flaton potential at the origin,
which is of order  $m^{2}M^{2}$.
With $M\sim 10^{12}$GeV and $m\sim10^{2}$GeV, this gives 
$V_{0}^{1/4}\sim 10^{7}$GeV which satisfies the condition
to avoid the excessive regeneration of light stable
fields.
Thermal inflation starts when the thermal energy
density falls below $V_{0}$ which corresponds to a 
temperature roughly of order $V_{0}^{1/4}$, 
and it ends when the
finite temperature becomes ineffective at a temperature 
of order $m$.
The number of e-folds is $N_{e}^{th}=\frac{1}{2}
ln(M/m)\sim 10$, which is
much smaller than the primordial inflation.
There is also the intriguing possibility that two or
more bouts of thermal inflation can occur in
succession, allowing more efficient solution of the
moduli problem.
In such cases  the number of e-folds will be
 about $N_{e}=10-25$\cite{thermal}.

\section{Unstable Domain Walls}
\hspace*{\parindent}
The discrete symmetry appears in many supersymmetric models.
The origins of such symmetries can be traced back to 
 $U(1)_{R}$ anomaly in the dynamical sector, or
the discrete symmetry in the superstring models 
which appears as the consequence of  possible 
compactification schemes.
In this section we make a collective review of such discrete
symmetries.
Here we consider a domain wall with the energy scale
$\sigma\sim \Lambda_{w}^{3}$.

\underline{R-changed}

In the supergravity mediated models for dynamical 
supersymmetry breaking, gaugino condensation in the
hidden sector of order $ 10^{10-12}$GeV 
is mediated by gravity, bulk fields or other interactions
such as an anomalous $U(1)_{X}$ field and
induces  soft supersymmetry breaking terms in the 
observable sector.
In the simplest hidden sector model (supersymmetric
Yang-Mills), 
$U(1)_{R}$ symmetry in 
the hidden sector is broken by anomaly and
a discrete R symmetry remains.
The discrete R symmetry is then spontaneously
broken when gaugino condensate.
This is a well-known example of the BPS domain 
wall\cite{Kovner_Shifman} in the global supersymmetric
gauge theory.
One may expect that the domain wall structure in the
hidden dynamical sector is a common feature of such
models for dynamical supersymmetry breaking.

On the other hand, in the gauge mediated models for
dynamical supersymmetry breaking, $U(1)_{R}$ symmetry
is strongly connected to the supersymmetry breaking.
The presence of an  $U(1)_{R}$ symmetry is a necessary 
condition
for supersymmetry breaking and a spontaneously broken
 $U(1)_{R}$ symmetry is a sufficient condition provided two
conditions are satisfied.
These conditions are genericity and caliculability.
This means that the domain wall structure in the dynamical
sector is not a common feature of the gauge mediated
supersymmetry breaking models.
However, there are many models which do not satisfy 
the genericity condition\cite{non-generic} 
where the  $U(1)_{R}$ symmetry will be anomalous.
In these models a discrete R symmetry is implemented 
and is spontaneously broken at relatively low energy scale 
of order $ 10^{5-9}$GeV.
In these models for dynamical supersymmetry breaking,
the dynamical sector may have a domain wall
configuration at intermediate scale 
$\Lambda_{w}\sim 10^{5-9}$GeV.

The R-charged domain wall configuration may also appear 
as the 
consequence of the 
spontaneous symmetry breaking of the explicit
$Z_{n}^{R}$ symmetry, which  is sometimes imposed by
 hand in order to solve phenomenological difficulties 
such as the $\mu$-problem\cite{Abel} or
the cosmological moduli problem\cite{Asaka}.
In these models the  scales of the domain walls are
determined by  phenomenological requirements,
 typically at $\Lambda_{w}\sim 10^{5- 10}$GeV.

When the universe undergoes a phase transition associated
 with  spontaneous breaking of such discrete symmetries,
 domain walls will inevitably form.
These domain walls are generally not favorable
 if they are stable.

In ref.\cite{matsuda_cosmo} 
we have shown that a constant term in the superpotential 
 always breaks the degeneracy of vacua  when supergravity is 
turned on, and that 
the pressure induced by the constant term satisfies
the usual condition for the safe decay of unwanted
domain walls.
In general, the constant term is required to make the
 cosmological 
constant very small, which is an inevitable feature of any 
phenomenological models for supergravity.
The magnitude of the energy difference $\epsilon$
induced by the
constant term is
$\epsilon\sim \sigma^{2}/M_{p}^{2}$
where $\sigma$ is the surface energy density of the wall.
This satisfies the usual condition for the safe decay of 
the cosmological domain wall.
Although the model is technically non-generic because it 
includes a single term which explicitly breaks the 
$Z_{n}^{R}$ symmetry, it is still a reasonable model 
for a vanishing cosmological constant.
In this sense,  the basic idea is similar to the 
well-known mechanism for the mass generation of the 
R-axion\cite{Axion}.

\underline{Not R-charged}

On the other hand, for the walls which do not
have R charge, the explicit symmetry breaking term should
be added by hand.
Since the gravity interaction does not  respect
 global symmetries such as the discrete symmetries we are
concerned about, the explicit breaking terms may appear as
 higher order terms suppressed by the Planck mass.
In this case, however, there is no reason
to expect that the magnitude of the energy difference
appears at their lowest bound
  $\epsilon\sim \sigma^{2}/M_{p}^{2}$.

\section{Unstable Domain Walls and Thermal Inflation}
\hspace*{\parindent}
In this section we shall discuss the formation, evolution
 and collapsing process of the cosmological domain walls
paying attention to the changes that should be
induced by  weak scale inflation.

In general, initial shape of the walls right after the
phase transition is determined by the random variation
of the scalar VEV.
One expects the walls to be very irregular, random
surfaces with a typical curvature radius, which
is determined by the correlation length of the
scalar field.
To characterize the system of domain walls,
one can use a simulation\cite{simulation}.
The system will be dominated by one large (infinite size)
wall network  and some finite closed walls (cells) when
they are produced.
The isolated closed walls smaller than the horizon
will shrink and disappear soon after the phase transition.
As a result, only a domain wall stretching across the
horizon will remain.
The initial distribution  of the cosmological domain 
walls  after the primordial inflation is not determined
solely from the thermal effect of reheating.
In some cases non-linear dynamics of the fields
(parametric resonance\cite{PR}, for example)
 will be important.
Here we do not discuss further on these topics
and temporarily make a simple assumption that the walls
 are produced just after the end of the primordial inflation.

For the walls $\Lambda_{w}<10^{11}$GeV and $\epsilon\sim
\sigma^{2}/M_{p}^{2}$, thermal inflation 
occurs before they collapse.
Such walls may experience a large though not huge number of
e-folds.
Extended structures arising from such weak scale
inflations are not necessarily inflated away.
Since the cells of the false vacuum cannot decay 
soon if they are much larger than the horizon 
scale\footnote{Here the  bubble nucleation rate is extremely 
small. We do not consider the false vacuum annihilation 
induced by the bubble nucleation, because such a scenario
is not realistic in our model.}, 
we expect that additional
 constraints should be required for such walls to decay.

When thermal inflation starts,
the initial scale of the domain wall network is 
the same as the particle horizon.
It  is of order $H_{th}^{-1}\sim (V_{0}^{1/2}/M_{p})^{-1}$,
where $H_{th}$ and $V_{0}$ denote the Hubble 
constant and the vacuum energy during thermal inflation.
The cells inflate during thermal inflation, then become
the scale of order $ l_{e}=H_{th}^{-1}e^{N_{th}}$.
Here $N_{th}$ demotes the number of
e-folds of thermal inflation.
Since  weak scale inflation may occur in succession
\cite{thermal},
 the number of e-folds $N_{th}$ will be the sum of 
these succeeding
weak scale inflations; $N_{th}\sim 10-25$.

At the end of weak scale inflation (at the time $t=t_{0}$),
\begin{equation}
\left(\frac{l}{d_{H}}\right)_{t=t_{0}}\sim
e^{N_{th}},
\end{equation}
where $d_{H}$ denotes the particle horizon.

After thermal inflation, coherent oscillation
of the inflation field $\phi$ starts.
The expansion during this epoch is estimated as\cite{text}
\begin{equation}
l=l_{t=t_{0}}\left(\frac{\rho_{\phi}}{V_{0}}
\right)^{-\frac{1}{3}}.
\end{equation}
The horizon size during preheating is estimated as
\begin{equation}
d_{H}\sim \left(\frac{\rho_{\phi}^{1/2}}{M_{p}}
\right)^{-1}.
\end{equation}
Thus the ratio becomes
\begin{equation}
\left(\frac{l}{d_{H}}\right)_{t_{rad}>t>t_{0}}\sim
e^{N_{th}}\left(\frac{\rho_{\phi}}{V_{0}}
\right)^{1/6}.
\end{equation}
Here $t_{rad}$ denotes the time when radiation
domination starts.
Thereafter $l$ grows like $T^{-1}$, where $T$ denotes
the temperature in the radiation dominated era.
Assuming the radiation-dominated expansion, the ratio
will be
\begin{equation}
\label{ratio}
\left(\frac{l}{d_{H}}\right)_{t>t_{rad}}\sim
\left(\frac{l}{d_{H}}\right)_{t=t_{rad}}
\left(\frac{T(t)}{T_{D}}
\right).
\end{equation}
Here $T_{D}$ denotes the reheating temperature after 
thermal inflation.

The domain wall network enters the particle horizon
at the time $t_{s}$  when the ratio becomes
unity.
Assuming that this occurs before radiation energy dominates
the universe,
$\rho_{\phi}$ at the time $t_{s}$ is 
\begin{equation}
\rho_{\phi}|_{t=t_{s}}\sim e^{-6N_{th}}V_{0}.
\end{equation}
In the case that the expansion during thermal inflation 
is about $10^{5}$ and $V_{0}^{1/4}$ is about 
$10^{7}$GeV,
$\rho_{\phi}|_{t=t_{s}}$ is estimated as 
$\rho_{\phi}|_{t=t_{s}}\sim 10^{-2}(GeV)^{4}$.
Here we assumed that the reheating temperature of the
thermal inflation is very low ($T_{D}<1GeV$) in order
to ensure sufficient entropy production.
As we have discussed in Sec.2, the walls that do not decay
until they dominate the universe must be excluded.
In this case, the walls that dominate the universe when 
they enter the horizon (i.e.,$\Lambda_{w}\ge 10^{6}$GeV)
are ruled out.
This bound seems very severe, since the walls below 
this scale should be further constrained by primordial
nucleosynthesis.
On the other hand, one may think that the walls of the
 scale $\Lambda_{w}
<10^{7}$ should be produced during thermal inflation
and suffer less expansion.
Including this effect, the above constraint can be relaxed
to allow the walls of the scale
$\Lambda_{w}\sim 10^{5-6}$GeV.

As a result, in many types of 
supersymmetric theories, walls of the intermediate
scale produced just after  primordial inflation 
cannot decay before they dominate
the energy density of the universe, even if they
satisfy the usual condition.
It must also be noted that this result does not
depend on the magnitudes of the explicit breaking terms.
If the magnitudes of the explicit breaking terms exceed
 the Vilenkin's lowest bound, the domination by the false
vacuum energy begins at earlier epoch.
As a result, the situation becomes worse for such larger
magnitudes of the explicit breaking terms.
Of course, there may be an exception that the explicit
 breaking terms are quite large so that the walls decay
 before weak scale inflation.
This may happen for the walls $\Lambda_{w}>V_{0}^{1/4}$, 
although the requirement becomes very severe. 
For example, when $\Lambda_{w}=10^{9}$GeV the required
value of the energy difference is about 
$\epsilon>10^{5}\sigma^{2}/M_{p}^{2}$,
which is more than 
$10^{5}$ times larger than the usual bound.

Another way to avoid the domain wall problem is to gauge
the discrete symmetry so that there is really one vacuum.
However, nontrivial anomaly cancellation conditions
must be satisfied.
Sometimes it requires fatal constraint on the components
of the model.

\section{Conclusions and Discussions}
\hspace*{\parindent}
In this paper we have studied  the decaying 
processes of unstable domain
walls and shown that the processes should be changed 
significantly if weak scale inflation takes place.
As a result, the usual condition for the safe decay of 
the cosmological domain wall must be changed.
For walls which can decay into  particles in the
standard model, the energy scales of the walls are strongly
restricted.

Although the above constraints looks severe,
there are other possibilities related to the 
structure formation; cosmology with ultra-light
pseudo-Nambu-Goldstone bosons\cite{soft_wall,bias}.
Contrary to the ordinary types of cosmological domain walls
which we have discussed above, expansion during
thermal inflation is a good news for such  scenarios.
Late decay of the soft domain walls can be realized
as a natural consequence of  thermal inflation,
and can contribute to the large scale structure formation.
We will study this topic in the next paper.

\section{Acknowledgment}
We wish to thank N.Sakai, N.Maru, Y.Sakamura and
Y.Chikira for many helpful discussions.

\end{document}